\documentclass{article}
\usepackage{graphicx} % Required for inserting images
\usepackage{amsmath,amsfonts,amssymb}
\usepackage{setspace}
\usepackage{tocloft}
\usepackage{tabularx}
\usepackage{hyperref}
\usepackage{float}
\usepackage{enumitem}
\usepackage{authblk}

\providecommand{\keywords}[1]
{
  \small	
  \textbf{Keywords: } #1
}

\title{Elliptic Curve Pairing Stealth Address Protocols}
\author[1]{Marija Mikić}
\author[2]{Mihajlo Srbakoski}
\affil[1]{Faculty of Mathematics, University of Belgrade}
\affil[2]{Faculty of Mathematics, University of Belgrade}
\date{}

\begin{document}

\maketitle
\begin{abstract}
Protecting the privacy of blockchain transactions is extremely important for users. Stealth address protocols (SAP) allow users to receive assets via stealth addresses that they do not associate with their stealth meta-addresses. SAP can be generated using different cryptographic approaches. DKSAP uses an elliptic curve multiplication and hashing of the resulting shared secret. Another approach is to use a elliptic curve pairing. This paper presents four SA protocols that use elliptic curve pairing as a cryptographic solution. ECPDKSAPs are pairing-based protocols that include viewing key and spending key, while ECPSKSAP is a pairing-based protocol that uses a single key with which spending and the viewing key are derived. We find that ECPDKSAPs give significantly better results than DKSAP with the view tag. The best results are achieved with Protocol 3 (Elliptic Curve Pairing Dual Key Stealth Address Protocol), which is Ethereum-friendly. ECPSKSAP is significantly slower, but it provides an interesting theoretical result as it uses only one private key.
\end{abstract}


\section{Introduction}

In public blockchains, all transactions are recorded transparently so that the transaction history of a specific pseudonymous user can be traced. Blockchains such as Bitcoin and Ethereum are open and publicly accessible. Third parties can view the data and possibly identify the participants in a particular transaction. There is therefore a great need to introduce private transactions on public blockchains such as Bitcoin \cite{textbook21} and Ethereum \cite{textbook6}. There are different ways to protect the transaction amount, sender and recipient address. The most important are ring signatures \cite{textbook17}, zero knowledge proofs \cite{textbook18} and stealth addresses \cite{textbook5}.

One notable application of stealth addresses is in the area of donations. Stealth addresses are also important in connection with salary payments. SAP (Stealth Address Protocol) can also be useful for smart contract wallets, public goods funding, decentralized finance (DeFi) systems, and the non-fungible token (NFT) landscape, expanding the potential application areas for these protocols.

Blockchains such as Zcash \cite{textbook9} and Monero \cite{textbook7} have some of the cryptographic methods at their core that enable transaction privacy. Popular programmable blockchains, particularly Ethereum, do not have inherent, comprehensive privacy protection at the core protocol level, so additional measures are required at the application layer to achieve a similar level of privacy.

The first stealth addresses were introduced in the Bitcoin ecosystem. Further improvements were then made. In 2013, Nicolas van Saberhagen described the Crypto\-No\-te protocol \cite{textbook20}, which used stealth addresses to improve the privacy of blockchain transactions. Stealth addresses were integrated into the Monero blockchain when it was introduced in 2014. With Monero, stealth addresses are inherently built into the core protocol. The programmable nature of the Ethereum blockchain facilitates the development of SAPs on top of it.

Stealth addresses allow users to generate a unique, one-time address for each transaction instead of relying on a static, publicly identified address.

We can use different versions of the stealth address protocol. Some of them, such as DKSAP \cite{textbook2, textbook5} (Dual-Key Stealth Address Protocol), are based on a cryptographic mechanism similar to the Diffie-Hellman method \cite{textbook1}, i.e.\ the private and public keys for the stealth address contain a shared secret that can be computed by both the sender and the recipient. Stealth addresses can be generated with another cryptographic method: elliptic curve pairing \cite{textbook15}. In the papers \cite{textbook3, textbook4} is presented SAP based on pairing. In the paper \cite{textbook3}, there is a gap that allows the sender and the person holding the viewing key to pair to obtain the private key of the stealth address. In the paper \cite{textbook4}, the private key of the stealth address is such that it can be computed by the sender without teamwork.

We propose four protocols based on elliptic curve pairing that overcome the above shortcomings: three ECPDKSAPs (Elliptic Curve Pairing Dual-Key Stealth Address Protocol) and one ECPSKSAP (Elliptic Curve Pairing Single-Key Stealth Address Protocol). The first ECPDKSAP is such that one private key can correspond to different public keys, and it is not Ethereum-friendly (it cannot be used on the Ethereum blockchain because the public key is calculated differently - Ethereum uses the Secp256k1 elliptic curve and the public key is obtained by multiplying the private key and the generator point of this curve). Protocol 2 is also ECPDKSAP, but with it one private key corresponds exactly to one public key. And this protocol is not Ethereum-friendly either. In Protocol 3, we have eliminated all the shortcomings of Protocol 1 and Protocol 2 and developed a protocol that is also more efficient than previous two. In Protocol 3, one private key corresponds exactly to one public key and this protocol can be used on the Ethereum blockchain (it is Ethereum-friendly). The ECPSKSAP is such that one private key can correspond to different public keys, and it is not Ethereum-friendly (uses a single key with which spending and the viewing key are derived). It is important to note that all three ECPDKSAPs provide more powerful results in terms of address computation time compared to the results of DKSAP in the paper \cite{textbook2}.

The paper consists of nine sections. The first section is the introduction. In the second section is an overview of the key literature on stealth addresses. Section 3 is short and introduces us to the types and properties of elliptic curve pairing. In the fourth section are discussed protocols that already exist and were the motivation for our work. Since our results focus on SAP protocols that use elliptic curve pairing, we addressed such protocols in the fourth section and pointed out the shortcomings of some currently existing protocols. Section 5 is devoted to the three protocols we propose: ECPDKSAPs, which use elliptic curve pairing and have two private keys. In this chapter, we have five subsections. First, we have described the general idea behind all three protocols and introduced the notation that we use in the rest of the paper. In the first subsection, we introduced concepts such as the viewing and spending keys. Then we have devoted each subsequent subsection to protocols 1, 2 and 3 that we mentioned in the previous paragraph. In the fifth subsection, we examine the various implementations of view tags and show their advantages and disadvantages. Section 6 is devoted to another protocol we propose, ECPSKSAP, which is also based on elliptic curve pairing, but with it, from one private key we get viewing and spending keys. Section 7 gives an overview of the security implications of our proposed protocols. In Section 8, we compare the efficiency of ECPDKSAPs and ECPSKSAP as well as the efficiency of ECPDKSAPs with different view tags. We also compare Protocol 3 and DKSAP in Section 8. Section 9 is the conclusion.

\section{Related work}

This section offers a overview of the key literature and a brief summary of the current state of research on stealth addresses.

Stealth address protocols have seen significant development since their introduction in 2011. The DKSAP, implemented in Monero, has been foundational, with numerous enhancements aimed at improving security and efficiency. Notable advancements include multi-key management to mitigate specific attacks, optimization techniques for faster parsing, and the use of elliptic curve pairing to reduce key storage. Recent works also focus on making stealth transactions indistinguishable from regular transactions, improving privacy in blockchain systems.

 In 2013 and 2014, Nicolas van Saberhagen \cite{textbook20} and Peter Todd \cite{textbook19} presented the basic stealth address protocols underlying the development of the Dual-Key Stealth Address Protocol (DKSAP), which was subsequently implemented in the Monero blockchain in 2014. DKSAP is the first stealth address protocol that contains a "viewing key", a key that allows the entity that possesses it to calculate the public key of the stealth address without knowing the private key of the stealth address.

DKSAP suffers from the problem of temporary key leakage, i.e.\ it requires the sender to publish an ephemeral public key. The stealth address protocol presented in \cite{textbook16} does not have this problem and allows the stealth address transaction to be indistinguishable from other transactions. This advantage comes at the cost of efficiency, as the recipient must parse all transactions, not just the stealth address transactions.

In the papers \cite{textbook3, textbook4} are proposed constructions of SAPs based on elliptic curve pairing. Such a construction enables, in paper \cite{textbook4}, the appearance of the first stealth address protocol where the spending key and viewing key are generated from one private key, thus using half a storage to store the private keys. This construct was the inspiration and motivation for ECPSKSAP, which is described in detail in Section 6.

The paper \cite{textbook2} provides a detailed overview of the development of stealth addresses, addressing potential challenges such as DoS (denial-of-service) attacks and de-anonymization, while also proposing strategies to mitigate these issues.

Due to its efficiency, DKSAP is the most widely used stealth address protocol, implemented in cryptocurrency systems such as Monero and Umbra \cite{textbook8}. With DKSAP, the recipient must search the blockchain for transactions until a matching transaction is found. The recipient then uses the temporary public key and their own private key to calculate and confirm whether they is the intended recipient. Figure 1 shows how DKSAP works, while \cite{textbook2} provides a detailed explanation.
 
\begin{figure}[H]
  \centering
    \includegraphics[scale=0.46]{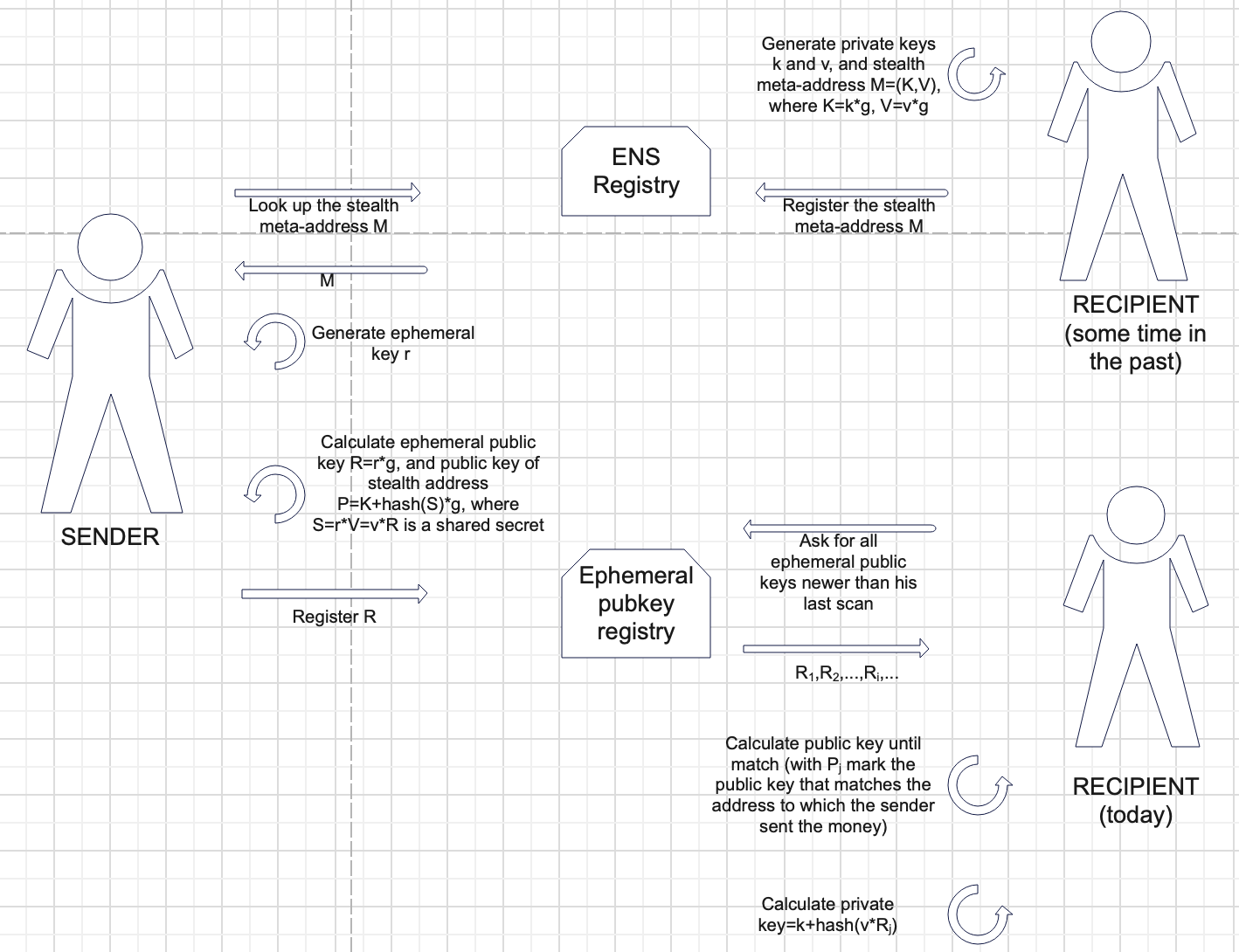}
      \caption{Dual-Key Stealth Address Protocol}
\end{figure}

In DKSAP, after extracting the ephemeral key, the recipient needs to perform two elliptic curve multiplications, two hash operations and one elliptic curve point addition, which limits the application of DKSAP in low-performance devices. One way to reduce the number of operations is to add a view tag to each ephemeral public key. View tag is a technique used in Monero \cite{textbook13} that allows recipients of stealth address transactions to skip certain steps in the parsing process under certain conditions. To achieve the desired goal, the view tag can be one byte of the hash(S) (this is what Monero uses). This way, the recipient only needs to perform a elliptic curve multiplication and a hash for each ephemeral public key. On average, the recipient would have to perform the remaining operations in only $\frac{1}{256}$ cases. In \cite{textbook2} the performance of DKSAP with and without the view tag is compared. Using the view tag reduces the parsing time for 5000 announcements by 86.96\%, for 10000 announcements by 86.92\%, for 20000 announcements by 86.96\%, for 40000 announcements by 86.97\% and for 80000 announcements by 86.99\%.

 BaseSAP \cite{textbook2} makes it possible to build various stealth address schemes on top of it. As shown in Figure 2, the implemented BaseSAP includes the basic components, including the announcer contract and the first two quads from left to right, which include the use of Secp256k1 with view tags and DKSAP. The remaining two quads show stealth address schemes that have piqued our interest - elliptic curve pairing-based and lattice-based schemes. Lattice-based cryptography, a key area of post-quantum cryptography, is a focus of our ongoing research. 

\begin{figure}[H]
    \centering
    \includegraphics[scale=0.13]{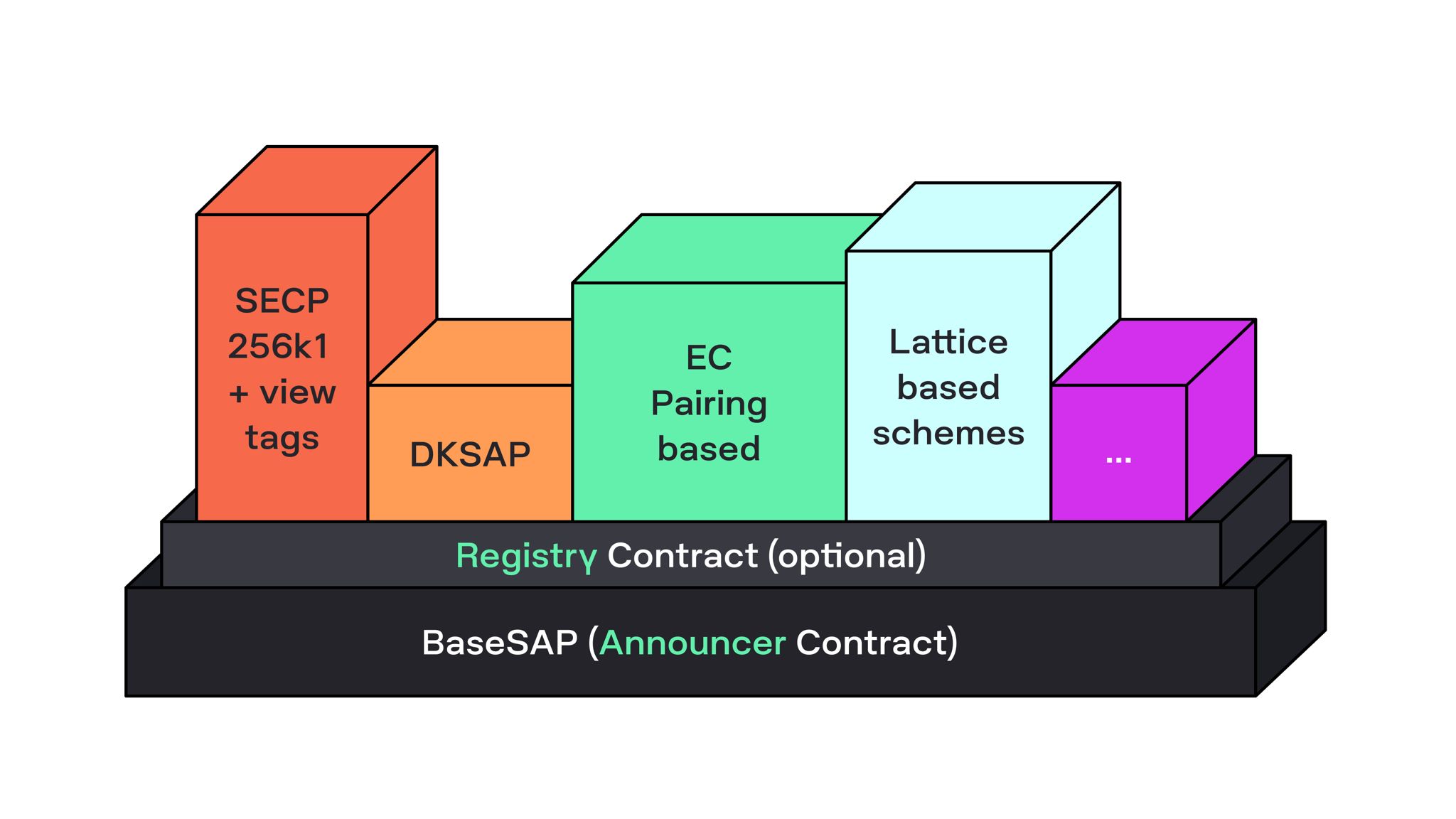}
    \caption{BaseSAP}
\end{figure}

\section{Elliptic curve pairing}

In this section we introduce the elliptic curve pairing, the operation on which ECPDKSAPs and ECPSKSAP are based.

\textbf{Definition:} Let $E$ be an elliptic curve over a finite field $K$. Let $\mathbb{G}_1$ and $\mathbb{G}_2$ be additively-written subgroups of order $p$, where $p$ is a prime number, of elliptic curve $E$, and let $g_1 \in \mathbb{G}_1, g_2 \in \mathbb{G}_2$ are the generators of groups $\mathbb{G}_1$ and $\mathbb{G}_2$ respectively. A map $e: \mathbb{G}_1 \times \mathbb{G}_2 \rightarrow \mathbb{G}_T$, where $\mathbb{G}_T$ is a multiplicatively-written subgroup of $K$ of order $p$, is called a elliptic curve pairing if satisfies the following conditions:
\begin{enumerate}
    \item $e(g_1, g_2) \neq 1,$ 
    
    \item $\forall R,S \in \mathbb{G}_1, \forall T \in \mathbb{G}_2:e(R+S,T) = e(R,T)*e(S,T),$ 

    \item $\forall R \in \mathbb{G}_1, \forall S,T \in \mathbb{G}_2:e(R,S+T) = e(R,S)*e(R,T).$ 
\end{enumerate}
The following properties of the elliptic curve pairing can be easily verified:
\begin{enumerate}
    \item $\forall S \in \mathbb{G}_1, \forall T \in \mathbb{G}_2:e(S,-T) = e(-S,T)=e(S,T)^{-1},$ 

    \item $\forall S \in \mathbb{G}_1, \forall T \in \mathbb{G}_2:e(a*S,b*T) = e(b*S,a*T)=e(S,T)^{a*b}.$ 
\end{enumerate}
In \cite{textbook14}, the following three basic types of pairing $e: \mathbb{G}_1 \times \mathbb{G}_2 \rightarrow \mathbb{G}_T $ are defined:
\begin{itemize}
    \item \textbf{Type 1:} $\mathbb{G}_1 = \mathbb{G}_2$, and we say $e$ is a symmetric pairing;
    \item \textbf{Type 2:} $\mathbb{G}_1 \neq \mathbb{G}_2$ and there exists an efficient homomorphism  $\phi: \mathbb{G}_2 \rightarrow \mathbb{G}_1 $, but no efficient one exists in the other direction;
    \item \textbf{Type 3:} $\mathbb{G}_1 \neq \mathbb{G}_2$ and there exists no efficient homomorphism between $\mathbb{G}_1$ and $\mathbb{G}_2$.
\end{itemize}

In our protocols we use type 3 elliptic curve pairing and their security is based on Bilinear Diffie-Helman Problem 1 (see \cite{textbook24}). 

\textbf{Definition:} Let $e: \mathbb{G}_1 \times \mathbb{G}_2 \rightarrow \mathbb{G}_T $ be a \textbf{type 3} elliptic curve pairing. Bilinear Diffie-Helman Problem 1 is to calculate $e(P,Q)^{ab},$ for given $P \in \mathbb{G}_1, aP \in \mathbb{G}_1, bP \in \mathbb{G}_1, Q \in \mathbb{G}_2$ and random $a,b$.

\section{Stealth address protocols based on pairing}

The protocols presented in the papers \cite{textbook3} and \cite{textbook4} are based on elliptic curve pairing, but both protocols have security vulnerabilities.

The protocol presented in the paper \cite{textbook3} uses \textbf{type 1} elliptic curve pairing $e: \mathbb{G} \times \mathbb{G} \rightarrow \mathbb{G}_T $.
A major security problem of this protocol is that when the sender gets the viewing key $v$, he can calculate the private key of the stealth address as follows:
$$
private \hspace{1mm} key=(k*v)*R=(k*v*r)*g=(r*v)*(k*g)=(r*v)*K,
$$
where $g$ is the generator of the group $\mathbb{G}$, $k$ is the recipient's private key, $K$ is the recipient's public key, where $K=k*g$, $r$ is the sender's ephemeral private key and $R$, where $R=r*g$, is the sender's ephemeral public key.
This scenario is possible when a person who has a viewing key contacts the sender and they agree to work together.

The protocol presented in the paper \cite{textbook4} uses \textbf{type 3} elliptic curve pairing $e: \mathbb{G}_1 \times \mathbb{G}_2 \rightarrow \mathbb{G}_T $ and requires only one private key instead of two, which saves storage space.
The problem with this protocol is that the private key of the stealth address can apparently be known to two parties:
$$private\hspace{1mm} key= e(R,k*g_2) = e(k*R,g_2) = e(r*K,g_2),$$
where $g_1$ and $g_2$ are the generators of the groups $\mathbb{G}_1$ and $\mathbb{G}_2$ respectively, $k$ is the recipient's private key, $K$ is the recipient's public key, where $K=k*g_1$, $r$ is the sender's ephemeral private key and $R$ is the sender's ephemeral public key, where $R=r*g_1$.
The private key of the stealth address is therefore known to both the sender and the recipient.

Although these two protocols are useless, they provide a good basis for the protocols presented in Section 5 and Section 6.

\section{ECPDKSAPs (Elliptic Curve Pairing Dual-Key Stealth Address Protocols)}

ECPDKSAPs are a pairing-based protocols that includes the viewing key and the spending key. We have implemented three different variants of the ECPDKSAP protocol:
\begin{itemize}
    \item Protocol 1: one recipient's private key corresponds to multiple stealth addresses and is not Ethereum-friendly, as the public key is calculated differently than on Ethereum, where the public key is obtained by multiplying the private key by the generator point of the elliptic curve Secp256k1;
    \item Protocol 2: one recipient's private key corresponds to exactly one stealth address and, like Protocol 1, is not Ethereum-friendly;
    \item Protocol 3: one recipient's private key corresponds to exactly one stealth address and it is Ethereum-friendly.
\end{itemize}

These protocols are developed in a progressive manner, with each successive version improving and refining the previous one by correcting its shortcomings. Protocol 3 is therefore the most advanced and widely applicable iteration, offering the greatest value.

In the context of ECPDKSAPs, any pairing-friendly elliptic curve and any pairing can be used. However, it has been shown that curve BN254 gives the best results, as shown in Section 8 using Protocol 3. The optimal Ate pairing has been found to be the best choice for pairing.
\subsection{Spending and viewing key}

The recipient of the transaction in the ECPDKSAPs generates two pairs of keys. The first pair of keys is the spending key $k$ and the corresponding public key $K$. The spending key is the private key of the stealth meta-address and is a number from the final field - the finite field associated with the elliptic curve. Although it is called the spending key, it is not actually the private key of the new stealth address, but it is crucial to obtain the private key of each recipient's stealth address. The public key corresponding to the spending key is obtained by multiplying $g_2$ by $k$, i.e.\ $K=k*g_2$ (in Protocol 1 and Protocol 2), where $g_2$ is the generator point of the subgroup $\mathbb{G}_2$ of the pairing-friendly elliptic curve and $k$ comes from $F_p$ - the finite field associated with the pairing-friendly elliptic curve, and in Protocol 3 the spending key is obtained by multiplying $g_e$ by $k$, i.e.\ $K=k*g_e$, where $g_e$ is the generator point of the elliptic curve Secp256k1 and $k$ comes from $F_{p_2}$ - the finite field associated with the elliptic curve Secp256k1.

Viewing key as a concept appeared later. It appeared for the first time in DKSAP in 2014. The recipient can give this key to someone if that person needs to be able to see all of recipient's stealth transactions (e.g.\ to a tax inspector). The person holding the viewing key cannot spend assets from stealth addresses, but can view transactions, because can calculate the public key of stealth addresses, but not the private key, which is necessary for spending assets. The viewing key $v$ is a private key, i.e.\ a number from the final field $F_p$. The public key $V$, which corresponds to the viewing key, results from the multiplication of $v$ and $g_1$, i.e.\ $V=v*g_1$, where $g_1$ is the generator point of the subgroup $\mathbb{G}_1$ of the pairing-friendly elliptic curve. In the ENS (Ethereum Name Service) registry, the recipient only adds the public keys associated with the spending and viewing keys, i.e.\ the points $K$ and $V$. ENS registry is a domain naming system built on the Ethereum blockchain. It presents an open, decentralized and extendable naming system as an alternative to the general operation of centralized domain naming services (DNS).

The viewing key of the stealth address is extremely important from a regulatory point of view, as it protects the privacy of the recipient of the transaction, but also offers the possibility of tracking all his transactions if necessary. This makes sense especially if decentralized identities and verifiable credentials come into play. Currently, it is possible to register multiple stealth meta-addresses in the ENS registry. However, if in the future only one ENS name can correspond to a single identity, then the person holding the viewing key could be sure to have insight into all stealth addresses associated with a particular stealth meta-address.

\subsection{Protocol 1}

In Protocol 1, the private key for each recipient's stealth address is the same, because it is determined as a product of the recipient's spending and viewing keys. Furthermore, Protocol 1 is not Ethereum-friendly because the public key of the stealth address is determined by elliptic curve pairing and not by multiplying the private key by the generator point of the elliptic curve Secp256k1.

Let $e: \mathbb{G}_1 \times \mathbb{G}_2 \rightarrow \mathbb{G}_T $ be a \textbf{type 3} elliptic curve pairing. Let $g_1$ be the generator point of the subgroup $\mathbb{G}_1$ and $g_2$ the generator point of the subgroup $\mathbb{G}_2$ of the pairing-friendly elliptic curve.

Figure 3 illustrates Protocol 1, which works as follows: 

$\textbf{1.} $ The recipient generates their spending key $k$ and viewing key $v$ and calculates their stealth meta-address $M=(K, V)$, where $K=k*g_2$ and $V=v*g_1$.

$\textbf{2.} $ The recipient adds an ENS record to register $M$ as the stealth meta-address for their ENS name.

$ \textbf{3.} $ The sender searches for recipient's stealth meta-address $M$ on ENS registry, using recipient's ENS name.

$ \textbf{4.}  $ The sender generates an ephemeral key $r$, which only the sender knows and uses only once (to generate this one stealth address). The sender also generates ephemeral public key $R=r*g_1$ and publishes it in the ephemeral public key registry. To optimize the recipient's search in the ephemeral public key registry, the sender sends the view tag (the view tag is explained in subsection 5.5) together with the public key $R$.

$ \textbf{5.}  $ The sender calculates the public key of the stealth address as $e(r*V, K)$. The sender can now send assets to this address. The public key of the stealth address can only be calculated by the sender and by the person who possesses the viewing key, since either the private ephemeral key $r$ (which only the sender possesses) or the viewing key $v$ is required (because $r*V=v*R$).

$ \textbf{6.}  $ The recipient searches through all public keys that have been published in the ephemeral public key registry since their last scan until finds a public key that matches the stealth address to which the sender sent the assets (for a quick search, use the view tag).

$ \textbf{7.}  $ The recipient's private key of the stealth address is $k*v$ and calculates the public key of the address as $e(R,g_2)^{private \hspace{0.6mm} key}$. The private key of the stealth address can only be calculated by the recipient, as requires the two private keys $k$ and $v$, which only the recipient possesses.

\begin{figure}[H]
  \centering
    \includegraphics[scale=0.39]{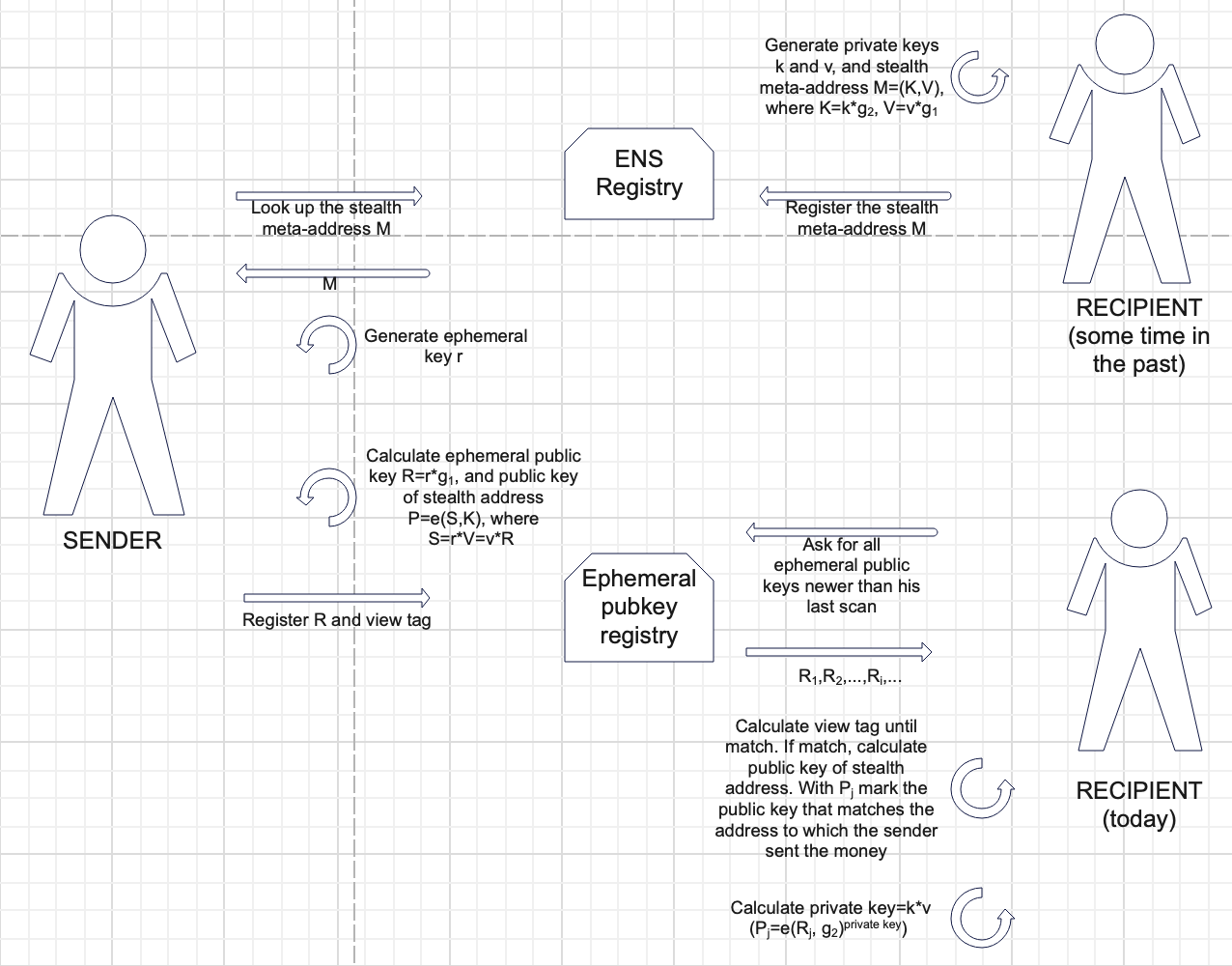}
    \caption{Elliptic Curve Pairing Dual-Key Stealth Address Protocol 1}
\end{figure}

In this protocol, after taking the ephemeral public key, the recipient must perform an elliptic curve multiplication and pairing to obtain the stealth address public key, and a hash operation to obtain the stealth address. This can be optimized by using view tag, which is explained in subsection 5.5.
The view tag optimization can also be applied to Protocols 2 and 3, which is also covered in subsection 5.5.

\subsection{Protocol 2}

Protocol 2 uses the same pairing as Protocol 1. The main difference is that the private key of the stealth address also depends on the ephemeral key, so the private key is different for each stealth address, which was not the case with Protocol 1. 

Protocol 2 is shown in Figure 4. The steps in Protocol 2 are as follows:

$\textbf{1-4.} $ Same as in Protocol 1.

$ \textbf{5.}  $ The sender calculates the public key of the stealth address as $e(\mathrm{hash}(r*V)*g_1, K)$. The sender can now send assets to this address. The public key of the stealth address can only be calculated by the sender and by the person who possesses the viewing key, since either the private ephemeral key $r$ (which only the sender possesses) or the viewing key $v$ is required (because $r*V=v*R$).

$ \textbf{6.}  $ Same as in Protocol 1.

$ \textbf{7.}  $ The recipient's private key of the stealth address is $k*\mathrm{hash}(v*R)$ and calculates the public key of the address as $e(g_1,g_2)^{private \hspace{0.6mm} key}$. The private key of the stealth address can only be calculated by the recipient, as requires the private key $k$, which only the recipient possesses.

\begin{figure}[H]
  \centering
    \includegraphics[scale=0.39]{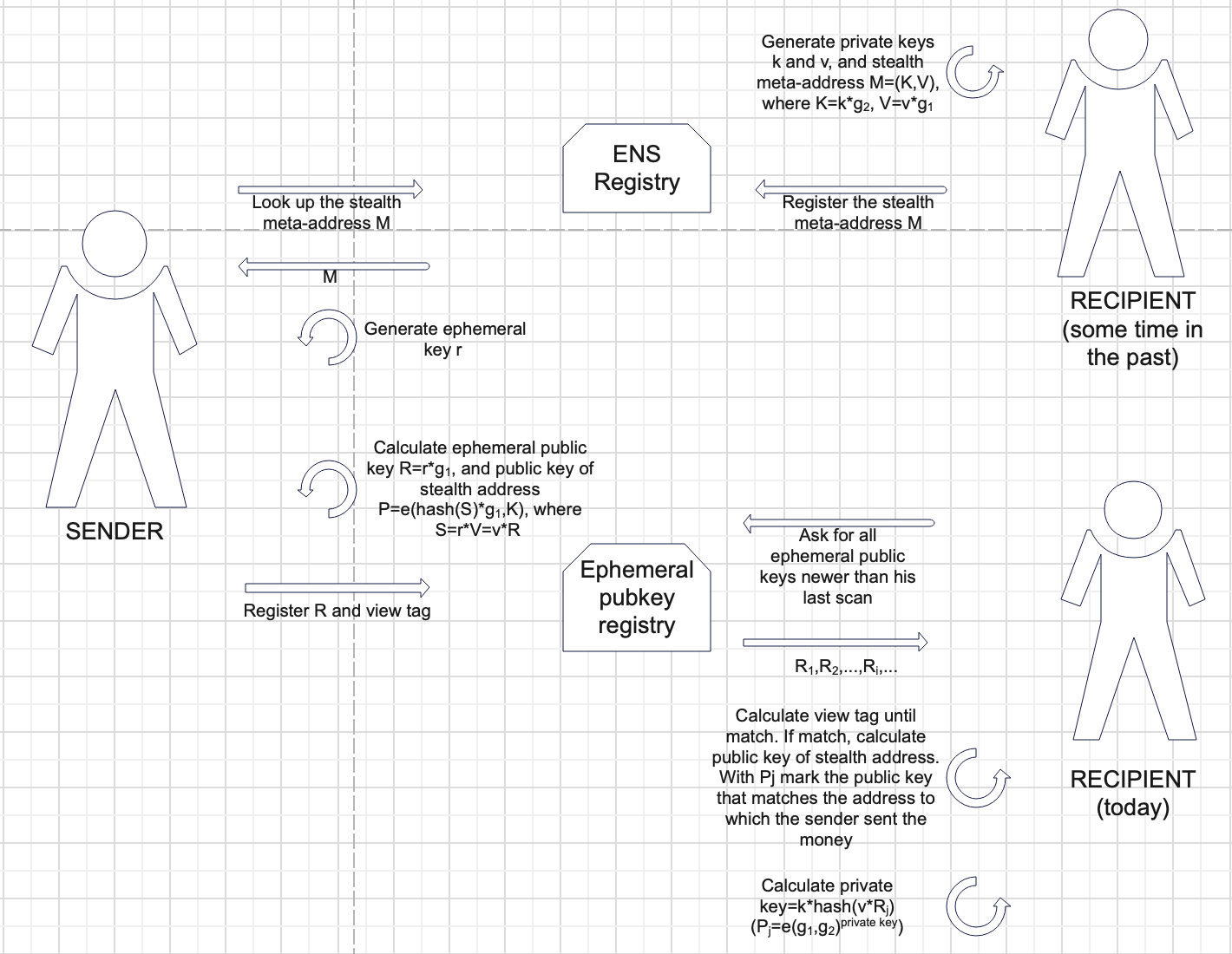}
    \caption{Elliptic Curve Pairing Dual-Key Stealth Address Protocol 2}
\end{figure}

In Protocol 2, after taking the ephemeral public key, the recipient must perform two elliptic curve multiplications, one pairing and two hash operations to obtain the stealth address. The operations required when using the view tag are explained in subsection 5.5.

\subsection{Protocol 3}

Protocol 3 is an Ethereum-friendly protocol because the public key of the stealth address is obtained by multiplying the private key with the generator point of the elliptic curve Secp256k1. Furthermore, each stealth address has a different private key, because the private key depends on the ephemeral key generated for each stealth address transaction.

Let $e: \mathbb{G}_1 \times \mathbb{G}_2 \rightarrow \mathbb{G}_T $ be a \textbf{type 3} elliptic curve pairing. Let $g_e$ be the generator point of the elliptic curve Secp256k1, and let $g_1$ and $g_2$ be the generator points of the subgroups $\mathbb{G}_1$ and $\mathbb{G}_2$ of the pairing-friendly elliptic curve.

Figure 5 illustrates Protocol 3, which works as follows:

$\textbf{1.} $ The recipient generates their spending key $k$ and viewing key $v$ and calculates their stealth meta-address $M=(K, V)$, where $K=k*g_e$ and $V=v*g_1$.

$\textbf{2-4.} $ Same as in Protocol 1.

$ \textbf{5.}  $ The sender calculates the public key of the stealth address as $b*K$, where $b$ is the first coordinate of $e(r*V,g_2)$. The sender can now send assets to this address. The public key of the stealth address can only be calculated by the sender and by the person who possesses the viewing key, since either the private ephemeral key $r$ (which only the sender possesses) or the viewing key $v$ is required (because $r*V=v*R$).

$ \textbf{6.}  $ Same as in Protocol 1.

$ \textbf{7.}  $ The recipient's private key of the stealth address is $b*k$ and calculates the public key of the address as $private \hspace{1mm} key*g_e$. The private key of the stealth address can only be calculated by the recipient, as requires the private key $k$, which only the recipient possesses.
\begin{figure}[H]
  \centering
    \includegraphics[scale=0.39]{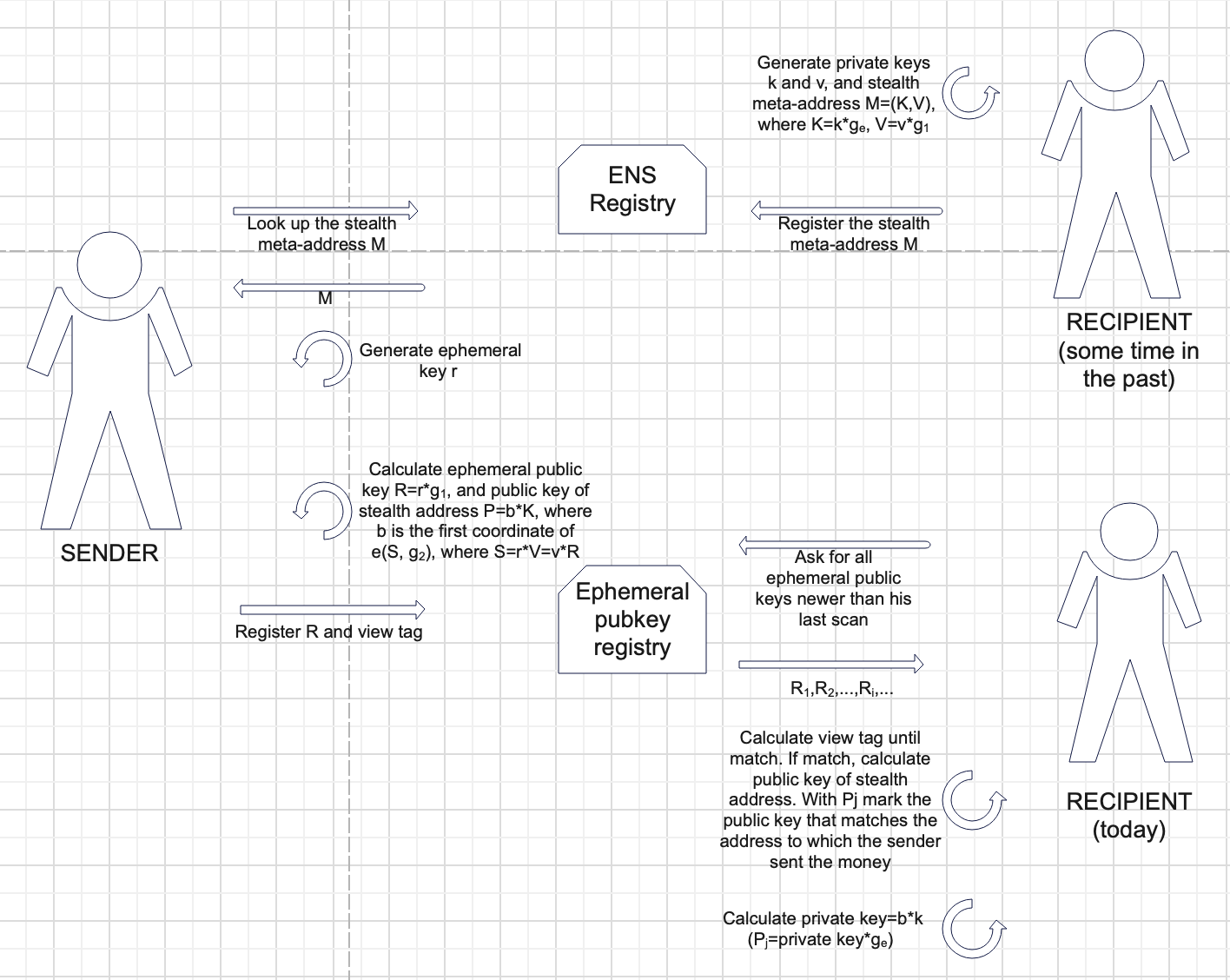}
    \caption{Elliptic Curve Pairing Dual-Key Stealth Address Protocol 3}
\end{figure}

After taking the ephemeral public key, the recipient must perform an elliptic curve multiplication, a pairing and a hash operation to obtain the stealth address. As in Protocols 2 and 3, this can be optimized by using view tag, which is explained in subsection 5.5.

\subsection{View tag}

The Monero blockchain was the first to start using view tag. Adding a view tag to the ephemeral public key is a technique to search the ephemeral public key registry more efficiently, so that not all steps need to be computed for each ephemeral public key, but only for the ephemeral public keys for which the view tag matches. The view tag can be added in one of two ways:
\begin{itemize}
 \item one (or more) byte(s) of the $x$-coordinate of the elliptic curve point $r*V$. Table 1 shows, for each ECPDKSAP, the operations that the recipient should perform when searching in the ephemeral public key registry. The operations are divided into those that must be performed for each announcement (i.e.\ operations to compute one (or more) byte(s) of the $x$-coordinate of the elliptic curve point $r*V$) and operations that are only performed after the view tag match. To calculate the view tag, an elliptic curve multiplication (ecMUL) must be performed. Pairing and hash should then be performed in all three protocols, while Protocol 2 requires an additional hash and elliptic curve multiplication.
 
\vspace{2mm}

\begin{table}[H]
    \centering
    \begin{tabular}{|l|c|c|}
    \hline
       & \textbf{operations before} & \textbf{operations after} \\
       \hline
        Protocol 1   & 1x ecMUL   &  1x pairing, 1x hash    \\
        \hline
        Protocol 2   &  1x ecMUL   &  1x ecMUL, 1x pairing, 2x hash    \\
        \hline
        Protocol 3   &  1x ecMUL   &  1x pairing, 1x hash    \\
        \hline
    \end{tabular}
    \caption{Operations before and after the view tag match.}
\label{table:1}
\end{table}

\item one (or more) byte(s) of the hash of the elliptic curve point $r*V$ (note that $\mathrm{hash}(r*V)$ does not have to be in the finite field $F_p$ for the calculation of the view tag). Table 2 shows, for each ECPDKSAP, the operations required to calculate this view tag as well as the operations that are only performed after the view tag match. The calculation of this view tag requires one more hash, while the necessary operations after the view tag match are the same as in the previous case, only we have one less hash in Protocol 2.

\vspace{2mm}

\begin{table}[H]
    \centering
  \begin{tabular}{|l|c|c|}
    \hline
       & \textbf{operations before} & \textbf{operations after} \\
       \hline
        Protocol 1   & 1x ecMUL, 1x hash     & 1x pairing, 1x hash     \\
        \hline
        Protocol 2   &  1x ecMUL, 1x hash   &   1x ecMUL, 1x pairing, 1x hash   \\
        \hline
        Protocol 3   & 1x ecMUL, 1x hash     & 1x pairing, 1x hash     \\
        \hline
    \end{tabular}
    \caption{Operations before and after the view tag match.}
\label{table:2}
\end{table}

\end{itemize}

If the size of the view tag is $n$ bytes, the expected number of the view tag matches until the corresponding ephemeral public key is found is $\frac{A}{256^n}$, where $A$ is the number of new ephemeral public keys since the last scan of the recipient. In Section 8, we present the time required by the recipient to search in the ephemeral public key registry, for different numbers of ephemeral public keys and different variants of the view tag for each ECPDKSAP protocol. Increasing the size of the view tag increases the efficiency of the protocol, but may lead to a reduction in security. Therefore, in Section 7, we analyse how each variant of the view tag affects the security of each ECPDKSAP.

\section{ECPSKSAP (Elliptic Curve Pairing Single-Key Stealth Address Protocol)}

ECPSKSAP is a memory-efficient stealth address protocol, as it requires half as much memory space to store the recipients' private keys compared to DKSAP and ECPDKSAPs. However, it is not directly compatible with Ethereum as it derives both the spending and viewing keys from a single key.

Let $e: \mathbb{G}_1 \times \mathbb{G}_2 \rightarrow \mathbb{G}_T $ be a \textbf{type 3} elliptic curve pairing. Let $g_1$ be the generator point of the subgroup $\mathbb{G}_1$ and $g_2$ the generator point of the subgroup $\mathbb{G}_2$ of the pairing-friendly elliptic curve. 

Figure 6 illustrates ECPSKSAP, which works as follows: 

$\textbf{1.} $ The recipient generates spending key $k$ and using this key calculates the viewing key $V$, where $V=k*g_2$. So, in this protocol, only the spending key $k$ is generated, while the viewing key is obtained from $k$. The stealth meta-address consists only of the public key $K$, where $K=k*g_1$.

$\textbf{2.} $ The recipient adds an ENS record to register $K$ as the stealth meta-address for their ENS name.

$ \textbf{3.} $ The sender searches for recipient's stealth meta-address $K$ on ENS registry, using recipient's ENS name.

$ \textbf{4.}  $ The sender generates an ephemeral key $r$, which only the sender knows and uses only once (to generate this one stealth address).

$ \textbf{5.}  $ The sender calculates the public key of the stealth address as $K+hash(S),$ where $S=e(K,g_2)^r$. The sender can now send assets to this address. The public key of the stealth address can only be calculated by the sender and by the person who possesses the viewing key, since either the private ephemeral key $r$ (which only the sender possesses) or the viewing key $V$ is required.

$ \textbf{6.}  $ The recipient searches through all public keys that have been published in the ephemeral public key registry since their last scan until finds a public key that matches the stealth address to which the sender sent the assets (for a quick search, use the view tag).

$ \textbf{7.}  $ The recipient's private key of the stealth address is $k+hash(e(R,V))$ and calculates the public key of the address as $private \hspace{0.6mm} key * g_1$. The private key of the stealth address can only be calculated by the recipient, as requires the two private keys $k$ and $V$, which only the recipient possesses.

In ECPSKSAP, after taking the ephemeral public key, the recipient must perform one elliptic curve pairing, two hash operations, one elliptic curve multiplication and one elliptic curve addition to calculate the stealth address. This can be optimized by using view tag so that the recipient only needs to perform one elliptic curve pairing and one hash operation for each ephemeral public key to calculate the shared secret, and only in $\frac{1}{256}$ (in the case of a one-byte view tag) would the recipient need to perform the remaining operations.

\begin{figure}[H]
  \centering
    \includegraphics[scale=0.39]{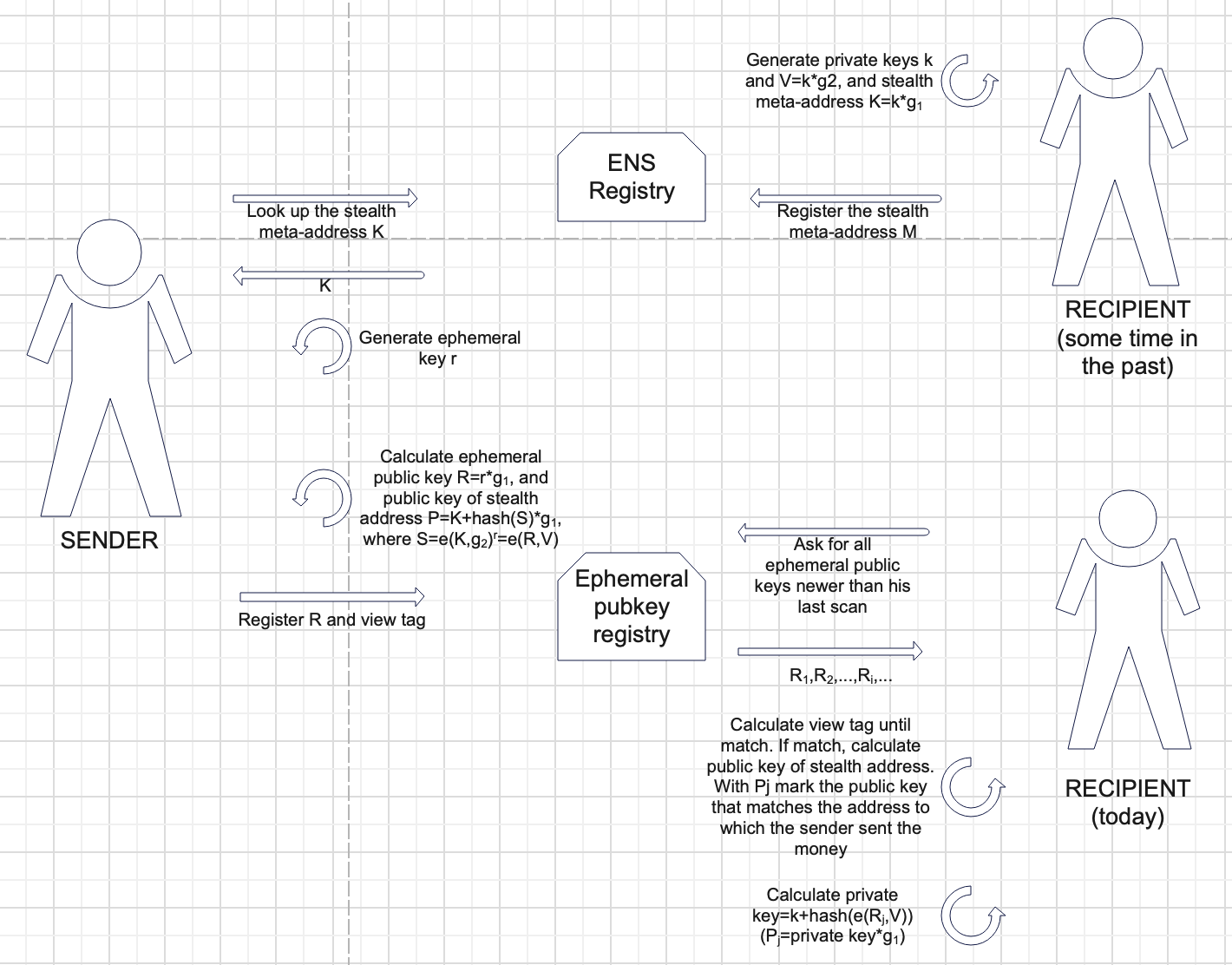}
    \caption{Elliptic Curve Pairing Single-Key Stealth Address Protocol}
\end{figure}

ECPSKSAP is significantly slower than ECPDKSAPs because a pairing is required to calculate the view tag, which is a much slower operation than the elliptic curve multiplication which is used in ECPDKSAPs to calculate the view tag.

\section{Security analysis}

The paper \cite{{textbook2}} presents a comprehensive analysis of the security implications of the DKSAP protocol, with a particular focus on denial-of-service (DoS) attack vectors and privacy concerns, especially the risk of de-anonymization of users. In DoS attacks, attackers attempt to overwhelm a system by sending an excessive number of requests, exhausting the system’s resources and rendering it unable to respond to legitimate users. Since the above paper addresses these concerns in detail, this paper focuses on the de-anonymization risks specific to recipients using the view tag mechanism.

De-anonymization in stealth address protocols refers to the risk that an attacker derives the public key $P$ of the recipient's stealth address. In the DKSAP protocol with the view tag, this vulnerability arises when an attacker is able to calculate the shared secret $S$ using the information from the view tag. Since the view tag is embedded in the hash of $S$, increasing the size of the view tag proportionally decreases the security level compared to the original 128-bit security provided by the Keccak-256 hash function. More specifically, the view tag of one byte reduces the security to 124 bits, with each additional byte reducing the security by a further 4 bits. It is therefore recommended to limit the size of view tags in DKSAP to a maximum of one byte to ensure the security of the protocol.

In subsection 5.5, two view tag variants are proposed for ECPDKSAPs:
\begin{itemize}
 \item one (or more) byte(s) of the $x$-coordinate of the elliptic curve point $r*V$;
\item one (or more) byte(s) of the hash of the elliptic curve point $r*V.$
\end{itemize}

In all three ECPDKSAPs, the elliptic curve point $r*V$ or its hash contributes to the computation of the public key of the stealth address, as it is the only element unknown to the attacker. Consequently, the disclosure of bytes of its $x$-coordinate via the view tag reduces the security of the protocol by 4 bits for each byte disclosed, starting from an initial security level of 128 bits. However, in Protocol 1 and 3, the view tag can be of any size by using bytes from $\mathrm{hash}(r*V)$ without compromising security. The reason for this is that $r*V$ is used directly for the calculation of the public key, rather than its hash. This approach provides a scalable solution that allows the view tag to grow in proportion to the increasing use of stealth addresses, thus increasing the efficiency of ephemeral public key searches. Protocol 2, like DKSAP, uses $\mathrm{hash}(r*V)$ to calculate the public key of the stealth address, so using the view tag reduces security.

To summarise, Protocol 3, which is Ethereum-friendly, allows the increase of the view tag without reducing security, which is not possible with the currently most commonly used DKSAP. Protocol 3 is therefore a more scalable solution than DKSAP, as the efficiency of the protocol can be increased if required by an increased volume of stealth address transactions.

\section{Implementation results and optimizations}

In our implementation of the ECPDKSAPs and ECPSKSAP protocols, we used the programming language \textit{Go} together with the library \textit{gnark-crypto} \cite{textbook26}. The full implementation of ECPDKSAPs is available in GitHub repository \cite{textbook25}.
 
 Since the most important point to optimize in stealth address protocols is the scanning of announcements (ephemeral public keys) in the ephemeral public key registry by the recipient, in subsections 8.1-8.4 we compared the protocols only for this part of the protocol. In subsections 8.1-8.3, scanning of announcement refers to the time it takes the recipient to compute the view tag for all announcements and perform additional operations to compute the public key, if the view tags match. In subsection 8.4, we also include the computation of the Ethereum address after the computation of the public key. Furthermore, subsection 8.5 shows the results of the individual operations that can be used to calculate the time of the entire protocol. The experiments were performed on an \textit{Apple MacBook Pro with M1 (10-core) chip and 16GB RAM}.

\subsection{Comparison of ECPDKSAPs and ECPSKSAP}

In this experiment, as in the experiments in subsections 8.2-8.4, to ensure consistency, measurements were made by randomly selecting 10 seeds used to generate private keys and announcements. The average time was used for the comparison.

Figure 7 shows a comparison of the scan times of the ephemeral public key registry with different numbers of announcements (5000, 10000, 20000, 40000 and 80000) between the ECPSKSAP protocol and the three variants of the ECPDKSAP protocol. In all cases, the view tag is set to one byte of $\mathrm{hash}(r*V)$. From a security perspective, as discussed in Section 7, it is better to use one (or more) byte(s) of $\mathrm{hash}(r*V)$ for the view tag in Protocol 1 and Protocol 3, rather than the $x$-coordinate $r*V$. The BN254 pairing-friendly elliptic curve and the optimal Ate pairing are used in all protocols. In Protocol 3, in addition to the BN254 elliptic curve, the Secp256k1 elliptic curve was used.

\begin{figure}[H]
  \centering
    \includegraphics[scale=0.37]{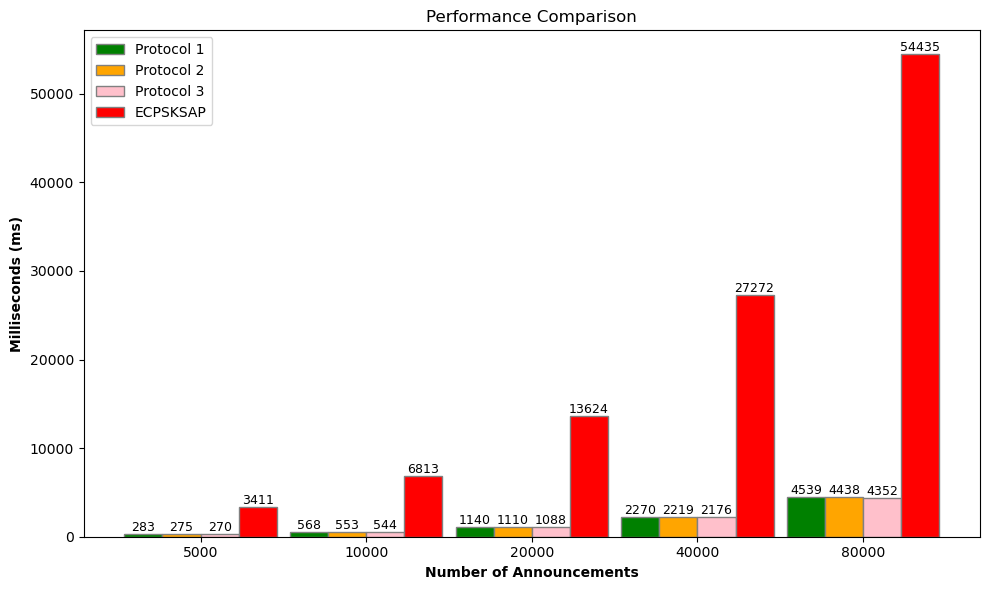}
      \caption{Timing test results for ECPDKSAPs and ECPSKSAP}
\end{figure}

Protocols 1, 2 and 3 show similar performance, with Protocol 3 being the fastest for the each number of announcements, while ECPSKSAP is significantly slower than the ECPDKSAP variants. Note that at 80000 announcements, Protocol 3 outperforms Protocol 1 by 4.12\% and Protocol 2 by 1.94\%.

\subsection{Comparison of view tags}

In subsection 8.1 we compared Protocol 1, Protocol 2 and Protocol 3 with the view tag of one byte of $\mathrm{hash}(r*V)$ and it is shown that Protocol 3 is the most efficient. Considering the security analysis done in Section 7, the number of bytes of $\mathrm{hash}(r*V)$ used as a view tag in Protocol 3 can be increased without compromising the security of the protocol. 

\begin{figure}[H]
  \centering
    \includegraphics[scale=0.37]{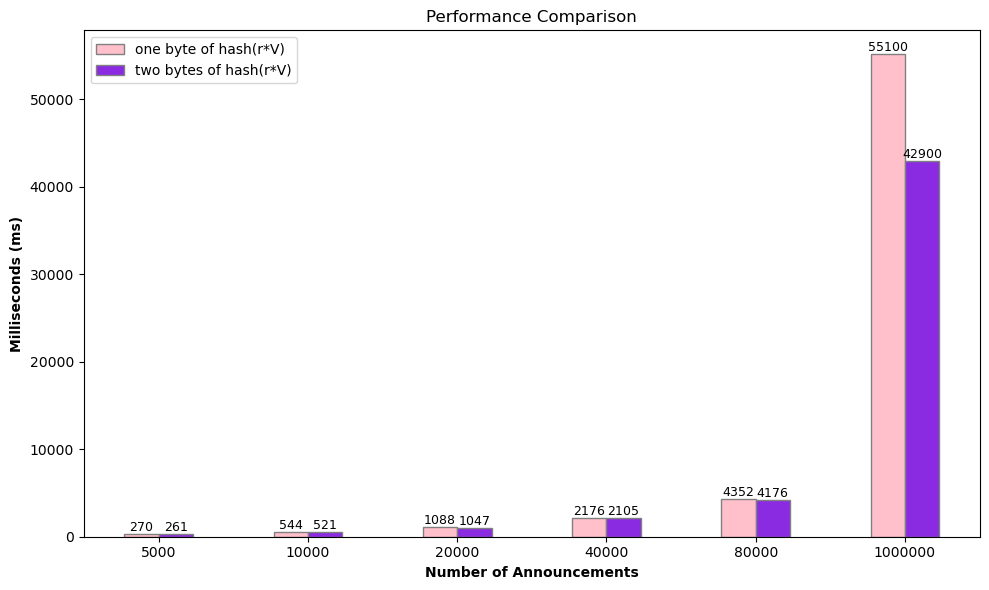}
      \caption{Protocol 3 - timing test results for different view tags}
\end{figure}

Therefore, we tested how much improvement of Protocol 3 can be achieved by increasing the view tag to two bytes of $\mathrm{hash}(r*V)$. The comparison of the scan times of the ephemeral public key registry with different numbers of announcements (5000, 10000, 20000, 40000, 80000 and 1000000) are shown in Figure 8.

Increasing the size of the view tag does not have a large impact on the efficiency of Protocol 3 when the number of announcements is small, but in the case of 1000000 announcements a significant improvement is achieved. Increasing the size of the view tag from one to two bytes of $\mathrm{hash}(r*V)$ reduces the scan time of the ephemeral public key registry from 55.1 seconds to 42.9 seconds, which is an improvement of 22.14\%.

\subsection{Comparison of elliptic curves}

Table 3 shows the results of scanning the ephemeral public key registry with 80000 announcements for different pairing-friendly elliptic curves. Protocol 3 with the optimal Ate pairing and the view tag of two bytes of $\mathrm{hash}(r*V)$ was used.

\begin{table}[H]
    \centering
    \begin{tabular}{|l|c|}
        \hline
        \textbf{Curve}  & \textbf{Time ($s$)} \\
        \hline
        BN254  & 4.18 \\
        \hline
        BLS12-377 &  7.11 \\
        \hline
        BLS12-381  & 7.08 \\
        \hline
        BLS24-315  & 5.54 \\
        \hline
        BW6-633  & 24.98 \\
        \hline
        BW6-761  & 38.11 \\
        \hline
    \end{tabular}
     \caption{Protocol 3 - timing test results for different elliptic curves}
\end{table}

The results show that the curve BN254 is the most optimal and that the curves from the families BLS12 and BLS24 give good results, while the curves from the family BW6 perform significantly worse.

\subsection{Comparison of Protocol 3 and DKSAP}

Protocol 3 and DKSAP, from the paper \cite{textbook2}, are both Ethereum-friendly and therefore represent solutions that use different cryptographic methods to generate stealth addresses on Ethereum.

Figure 9 shows a comparison of the scan times of the ephemeral public key registry for different numbers of announcements (5000, 10000, 20000, 40000 and 80000) of our implementation of Protocol 3 (with Secp256k1 elliptic curve, pairing-friendly curve BN254 and optimal Ate pairing) and the implementation of DKSAP from the paper \cite{textbook2}. The values of the private keys and the size of the view tag (5.5 bytes) from \cite{textbook2} were used.

The implementation of Protocol 3 shows a significant improvement compared to the mentioned implementation of the DKSAP protocol. For each number of announcements (5000, 10000, 20000, 40000 and 80000), the scan time of the ephemeral public key registry was reduced about 5 times, specifically in the case of 80000 announcements we achieved an improvement of 79.75\%.

\begin{figure}[H]
  \centering
    \includegraphics[scale=0.39]{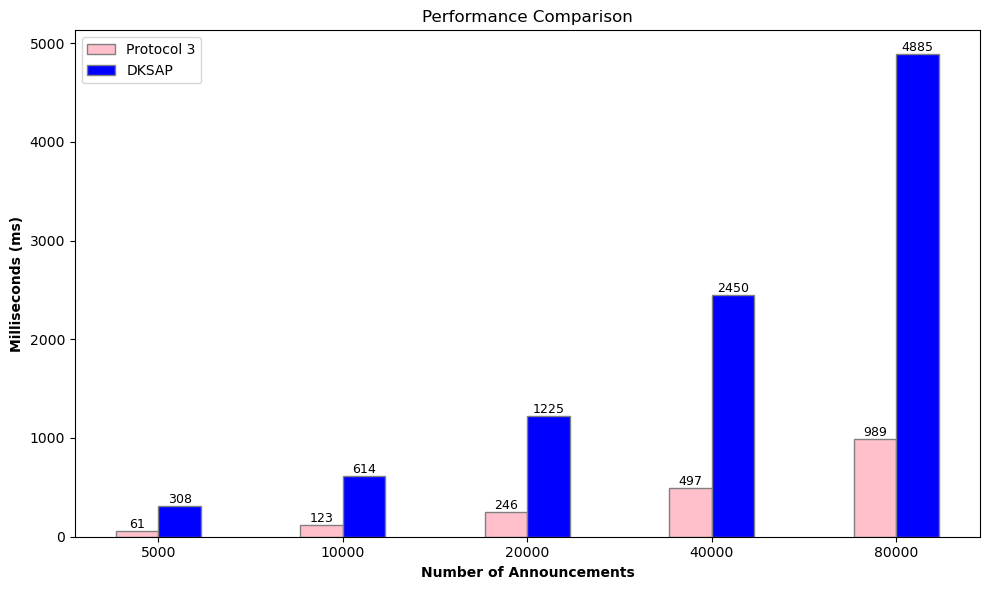}
      \caption{Timing test results for Protocol 3 and DKSAP}
\end{figure}

\subsection{Comparison of individual operations}

Table 4 shows the execution times for different cryptographic operations required to scan the ephemeral public key registry in Protocol 3. To ensure the accuracy and consistency of the measurements, each operation was repeated 1000 times and the average execution time is shown in the Table 4. In this evaluation, we tested an implementation of Protocol 3 that uses the Secpt256k1 elliptic curve, the BN254 pairing-friendly curve, the Keccak-256 hash function and optimal Ate pairing.

In our implementation, the elliptic curve multiplication and the pairing operations during the scanning of the ephemeral public key registry are divided into two stages: precomputation (executed once by recipient) and the remaining computations (executed for each ephemeral public key). This approach resulted in a 4.53\% improvement in elliptic curve multiplication and a 13.1\% improvement in pairing compared to using the \textit{gnark-crypto} implementation of these functions (\textit{ScalarMultiplication} and \textit{PairFixedQ}).

\begin{table}[H]
    \centering
    \begin{tabular}{|l|c|}
        \hline
        \textbf{Operation}  & \textbf{Time ($\mu s$)} \\
        \hline
        ecMUL precomputation  &  7.94 \\
        \hline
        ecMUL  &  56.5 \\
        \hline
        optimal Ate pairing precomputation  &  229 \\
        \hline
        optimal Ate pairing & 597   \\
        \hline
        Keccak-256 hash  & 2.33  \\
        \hline
    \end{tabular}
     \caption{Timing test results for individual operations}
\end{table}

Table 4 shows that pairing is the most timeconsuming operation. However, since pairing is only performed after the view tags match, it does not have a large impact on the effectiveness of Protocol 3.

Compared to the results from the paper \cite{textbook2}, by using the \textit{gnark-crypto} library we achieved an improvement of 92.11\% for the elliptic curve multiplication and an improvement of 27.39\% for the Keccak-256 hash.

\section{Conclusion}
Our main goal in this paper is to hide the connection between the recipient and their address in order to protect privacy and make this as efficient as possible. In this paper, we first gave a description of stealth address technique and introduced the most commonly used protocol at the moment - DKSAP (Dual-Key Stealth Address Protocol). We introduce four new protocols based on elliptic curve pairing: three ECPDKSAPs (Elliptic Curve Pairing Dual-Key Stealth Address Protocols) and ECPSKSAP (Elliptic Curve Pairing Single-Key Stealth Address Protocol). In the paper, we have shown how the privacy of the recipient of the transaction can be ensured using elliptic curve pairing. 

A security analysis was also performed with respect to the view tag. In addition, all our protocols were compared in terms of their efficiency (the time needed for the recipient of the transaction to compute the public key of their new stealth address), but also with DKSAP \cite{textbook2}. The obtained results clearly show that ECPDKSAP Protocol 3 is the most effective compared to our other protocols and DKSAP. We especially emphasize that we need about 5 times less time to compute the recipient's address compared to DKSAP \cite{textbook2}. Protocol 3 is Ethereum-friendly and open source and we believe that its implementation will be important from an application perspective (especially because of its efficiency).

\end{document}